# Influence of Plasmonic Array Geometry on Energy Transfer from a Quantum Well to a Quantum Dot Layer

*Luke. J. Higgins[1*], Cristian A. Marocico[1], Vasilios D. Karanikolas[1], Alan P. Bell[1], John J. Gough[1], Graham P. Murphy[1], Peter J. Parbrook[2] and A. Louise Bradley[1].*

[1]School of Physics and CRANN, Trinity College Dublin, College Green, Dublin 2, Ireland.

[2]Tyndall National Institute and School of Engineering, University College Cork, Lee Maltings, Prospect Row, Cork, Ireland.

Email Address: higginlu@tcd.ie

## ABSTRACT

A range of seven different Ag plasmonic arrays formed using nanostructures of varying shape, size and gap were fabricated using helium-ion lithography (HIL) on an InGaN/GaN quantum well (QW) substrate. The influence of the array geometry on plasmon-enhanced Förster resonance energy transfer (FRET) from a single InGaN QW to a ~ 80 nm layer of CdSe/ZnS quantum dots (QDs) embedded in a poly(methyl methacrylate) (PMMA) matrix is investigated. It is shown that the energy transfer efficiency is strongly dependent on the array properties and an efficiency of ~ 51% is observed for a nanoring array. There were no signatures of FRET in the absence of the arrays. The QD acceptor layer emission is highly sensitive to the array geometry. A model was developed to confirm that the increase in the QD emission on the QW substrate compared with a GaN substrate can be attributed solely to plasmon-enhanced FRET. The individual contributions of direct enhancement of the QD layer emission by the array and the plasmon-enhanced FRET are separated out, with the QD emission described by the product of an array emission factor and an energy transfer factor. It is shown that while the nanoring geometry results in an energy transfer factor of ~ 1.7 the competing quenching by the array, with an array emission factor of ~ 0.7, results in only an overall gain of ~ 14% in the QD emission. The QD emission was enhanced by ~ 71% for a nanobox array, resulting from the combination of a more modest energy transfer factor of 1.2 coupled with an array emission factor of ~ 1.4.

# INTRODUCTION

In recent times there have been numerous reports on the use of plasmonic structures, based on both metal nanoparticles and films, to enhance the emission of different emitters including semiconductor quantum wells (QWs) [1-6] and quantum dots (QDs).[7-12] The enhanced emission can arise due to increased absorption under optical pumping, increased radiative recombination rates and/or scattering by the plasmonic nanostructures themselves.[12] Along with coupling to a single emitter, plasmonic nanoparticles (NPs) can also be used to couple light between two emitters. In the 1980s it was theoretically proposed that localised surface plasmons (LSPs) could modify Förster resonance energy transfer (FRET) between two molecules.[13, 14] The first report of FRET was observed from smaller to larger CdSe core-only QDs by Kagan *et al.*[15, 16] Crooker *et al.* demonstrated that FRET could be optimized by arranging QDs of different sizes to create an energy gradient.[17] FRET has also been observed in several other QD hybrid material systems, including in QD-QD,[17-19] QD-dye,[20, 21] QD-graphene,[22] QD-QW [23] and QD-protein systems. The distance dependence of FRET is well-defined by a dipole-dipole interaction,[25, 26] typically occurring over short length scales of ~ 10 nm. However, by utilising the enhanced local fields generated around plasmonic NPs this distance can be substantially increased. The enhancement of the FRET process strongly depends on the distance from the metal NP, the shape of the metal NP and orientation of the dipoles with respect to the metal NP. Plasmon enhanced FRET between QDs was first experimentally observed in mixed monolayers[27] and enhancements of the energy transfer rate, efficiency, and range have been reported in a variety of material systems and geometries over the last decade.[10, 11, 27-40] Extensive studies on LSP mediated FRET revealed how it was influenced by factors such as the donor emission wavelength and the quantum yield, concentration of the NPs and relative positions of the donor and acceptor with respect to the nanoparticles.[32, 41-43]

Combining QWs and QDs has, to date, played a significant role in many devices for both white light generation[44-46] and colour conversion.[47-50] Radiative pumping of QDs by QWs, which depends on the QDs absorbing photons emitted by the QW into the far-field, is a multi-step process that is inherently lossy, for example due to the challenge of efficient photon extraction from an LED. Different strategies to efficiently extract photons from LEDs have been a topic of considerable research, including but not limited to modification of the surface roughness of the LED[51, 52] and resonant cavity structures.[62] However, near-field nonradiative energy transfer allows for the direct transfer of energy from the QW to the QDs, and is an alternative pumping mechanism. Basko *et al.* theoretically predicted nonradiative energy transfer or FRET from a QW to organic molecules.[53] However, it was Achermann *et al.* who experimentally demonstrated that efficient pumping through FRET from an InGaN/GaN QW to a monolayer of QDs could be achieved, and showed that this process is fast enough to compete with both radiative and nonradiative relaxation processes within the QW itself.[23, 48] Due to the strong distance dependence of FRET, the structure required a very thin QW capping layer and a single monolayer of QDs, resulting in a donor-acceptor separation of 7 nm. Subsequent studies have focussed on optimising geometries for FRET, such as dry etched nanopillars to bring a higher number of acceptor QDs close enough to benefit from nonradiative energy transfer from the QW,[54, 55] graded QDs[56] or optimized donor-acceptor ratios.[57] While the device performance benefitted from the proximity of the QDs to the QW, other drawbacks such as fabrication complexity and introduction of defects or the need to carefully position emitters have to be considered.

Incorporating plasmonic structures into a system consisting of a QW donor and QD acceptors has been shown to increase the energy transfer efficiency and to yield higher emission from a thicker QD layer. The plasmon-enhanced FRET efficiency is determined by the interplay between the QW spontaneous emission rate, the rate of nonradiative energy transfer to the plasmonic structure and the rate of plasmon-coupled nonradiative energy transfer to the acceptors. The plasmon-enhanced FRET must compete with the nonradiative energy transfer from the QW to the plasmonic structure. The influence of the plasmonic element on the interplay of these processes is complex and strongly dependent on the properties of the plasmonic component and how it couples to the donor and acceptor. Plasmon-enhanced FRET is sensitive not only to the spectral overlap of the donor emission with the acceptor absorption, but also with the spectrum of the plasmon mode.[10, 31] The optical properties of plasmonic NPs can be tuned by varying their material, size and shape.[58] This allows them to interact more strongly with different emitters by tuning the level of spectral overlap between the two species and also by changing the local field enhancement and the field distribution. By using plasmonic arrays these optical properties can be controlled even further by tuning the level of interaction between individual NPs, achieved by varying the gap between consecutive plasmonic elements. Work done on plasmon-enhanced FRET within different geometries showed that by changing the size of colloidal NPs, for example, the spectral overlap with the donor is automatically lost.[27, 31,35-37, 59] However, an array geometry provides sufficient flexibility to vary the individual nanoparticle shape while maintaining spectral overlap. Variation in the plasmonic array design can modify the level of LSP enhanced absorption and emission processes within the QW-QD system by changing the absorption and scattering profile, and will also impact the plasmon-enhanced FRET within the system.

In this paper a range of Ag plasmonic arrays with varying shape and size were fabricated using helium-ion lithography (HIL) on both a bulk GaN and an InGaN/GaN QW substrate to investigate plasmon-enhanced FRET from a single QW to a ~ 80 nm layer of CdSe/ZnS QDs dispersed in a poly(methyl methacrylate) (PMMA) matrix. The thickness of the QD acceptor layer and the dimensions of the Ag plasmonic NPs are large compared with previous reports of plasmon-enhanced FRET.[31, 36] Within the range of plasmonic arrays presented, different levels of both LSP enhanced FRET and enhanced absorption and scattering processes can be expected. The interplay between the processes will be elucidated. The ability to tune these processes could be exploited in applications for colour tuning or colour switching plasmonic devices. Colour switching could be achieved by selecting a plasmonic array that quenches the donor emission strongly while maintaining LSP-FRET to the acceptor. The colour tuning could be achieved by a combination of both LSP enhanced absorption and emission of either the donor or acceptor and enhanced FRET between them.

# EXPERIMENTAL METHODS

A schematic of the complete structure is shown in Figure 1 (a). Ag nanoparticle (NP) arrays were fabricated on a single InGaN QW structure and, subsequently, covered in a layer of CdSe/ZnS core-shell QDs embedded in a ~ 80 nm thick layer of PMMA. The single QW was grown by metal-organic vapour phase epitaxy (MOVPE) on a thick GaN layer on a c-plane sapphire substrate. The QW is 2 nm thick InGaN capped with a 3 nm GaN barrier. The photoluminescence (PL) spectrum of the QW, given in Figure 2 (a,b,c), shows a peak emission wavelength of 516 nm and a FWHM of 40 nm. The barrier thickness of 3 nm was selected to allow for strong near-field coupling between the excitons in the QW and the plasmonic NPs.

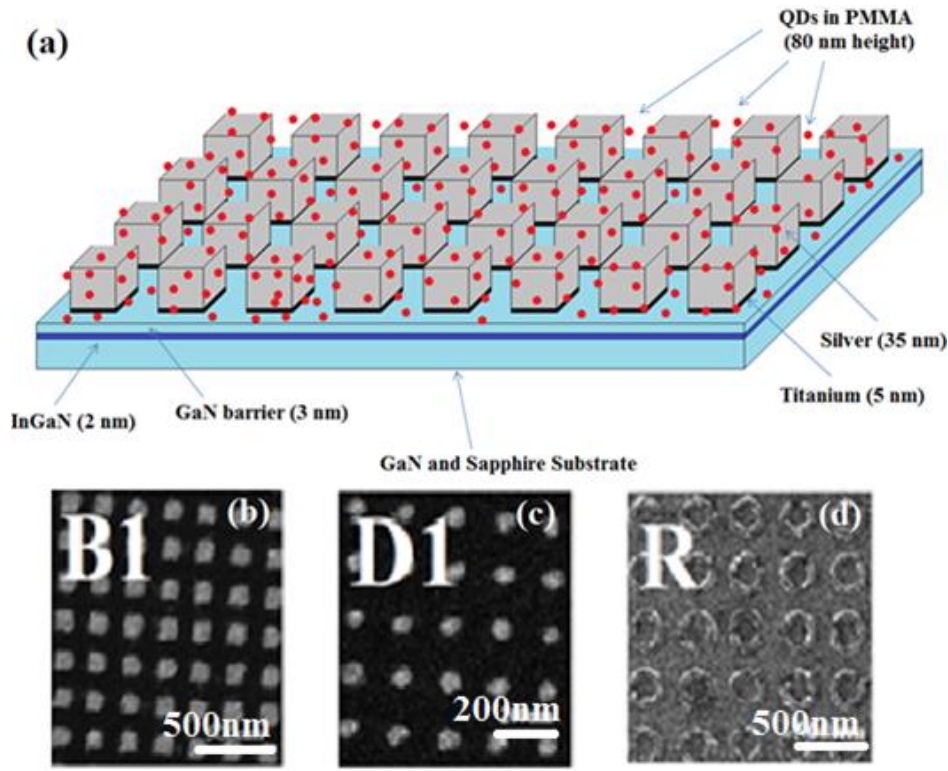

Figure 1: (a) Schematic of complete hybrid structure along with HIM images of a nanobox array (B1), nanodisc array (D1) and nanoring (R).

The strength of interaction could be controlled via the properties of the Ag NP arrays, which were fabricated using Helium-Ion Lithography (HIL). To fabricate the plasmonic arrays PMMA is first spun onto the GaN capping layer surface and patterned using a Zeiss Orion Plus helium-ion microscope. Following the development, 5 nm of Ti and 35 nm of Ag are deposited using a metal evaporation process. Following the lift-off process the plasmonic arrays are left on the surface. Each array contained 200 x 200 units arranged in a symmetric two-dimensional square lattice giving overall array dimensions of between 34 μm x 34 μm to 70 μm x 70 μm. The different nanostructures and dimensions are given in Table 1 and helium-ion microscopy (HIM) images for nanobox array B1, nanodisk array D1 and ring array R are shown in Figure 1 (b,c,d). HIM images for all seven arrays discussed in this paper are shown in the Supporting Information, Figure S1. The images clearly show that there is a large degree of uniformity across the arrays and that the aspect ratio of each individual unit is also well maintained.

| Label | Shape | Unit Dimension (nm) | Gap (nm) |
|---|---|---|---|
| B1 | Box | 100 | 100 |
| B2 | Box | 100 | 130 |
| B3 | Box | 100 | 160 |
| R | Ring | 250 Outer Diameter | |
| | | 150 Inner Diameter | 100 |
| D1 | Disc | 70 Diameter | 100 |
| D2 | Disc | 70 Diameter | 130 |
| D3 | Disc | 70 Diameter | 160 |

Table 1: Shapes and sizes of nanostructures fabricated directly onto QW and bulk GaN substrates.

The QDs are CdSe/ZnS core-shell nanocrystals with a diameter of 6.3 nm and a peak emission wavelength of 649 nm, seen in Figure 2 (a,b,c). The QDs are CdSe/ZnS core-shell nanocrystals with a diameter of 6.3 nm and a peak emission wavelength of 649 nm. The QDs layer was spin cast after adding the QDs in toluene to a 2.5 % wt. PMMA (also dissolved in toluene) solution. The resulting layer had a QD concentration of $(1.9 \pm 0.1) \times 10^{21}$ m$^{-3}$. The QD layer thickness was measured with a Dektac6 profilometer. It was found that the thickness of the PMMA layer is increased by $(19 \pm 5)$ nm on the array compared with a $(60 \pm 5)$ nm thickness off the array. The same solution of QDs is spun down on Ag array decorated GaN substrate for reference measurements.

The emission and energy transfer (ET) processes were characterised using PL spectroscopy and time-resolved photoluminescence (TRPL). All measurements were made at room temperature. For PL measurements the samples were excited at normal incidence using a pulsed 405 nm laser diode focused using a x40 objective lens. The PL spectra were recorded using a fiber-coupled Andor grating spectrometer. The PL decays were measured using a PicoQuant

Microtime 200 time-resolved confocal microscope with 150 ps temporal resolution. The QW or QD emission bands were selected using broad-band filters with a full-width-at-half-maximum of (70 ± 5) nm centred at 500 nm and 650 nm, respectively. The PL decays and spectra were recorded over an area of 20 μm × 20 μm. The PL decay curves for both the QW and the QDs are fit with a bi-exponential and the average PL lifetime is quoted, see Supporting Information, Equation (S1).

The NP arrays were designed using commercial finite difference time domain (FDTD) software. The absorption and scattering spectra and electric field intensity maps of the Ag NP array are simulated using plane-wave excitation. The data for arrays B1, D1 and R are shown in Figure 2, with the spectra for all arrays given in the Supporting Information, Figure S2. The simulations take account of the PMMA layer on top of the Ag nanoboxes and the 5 nm Ti layer. The dielectric permittivity of the Ag and Ti materials are included using experimentally measured data.[60] The GaN and PMMA layers are modelled with a constant dielectric permittivity of $\varepsilon_{GaN}$ = 5.35 and $\varepsilon_{PMMA}$ = 2.2, respectively. The extinction cross-section is given by $C_{Ext} = C_{Abs} + C_{Sca}$, where $C_{Abs}$ is the absorption cross-section and $C_{Sca}$ is the scattering cross-section. It can be seen from Figure S2 that for the box and nanoring arrays, a peak in the absorption spectra overlaps the donor emission spectrum in each case. For the nanodisc arrays there is overlap between the absorption spectra and the QW emission spectrum even though no clear peak in the array absorption spectra is present. The level of scattering at the acceptor wavelength clearly varies from one array to another and this will impact the degree of interaction between the QDs and the arrays by themselves on the bulk GaN substrate. Field intensity maps for all structures in the x-z plane, at excitation wavelengths corresponding to the donor QW and acceptor QD peak emission, are given in the Supporting Information, Figure S3 and Figure S4. Variation in both the magnitude and extent of the electric field intensity across the structures can be seen, showing strong localization to the nanobox, while extending for a great distance through the centre of the nanoring. The spatial extent of the field is expected to influence the volume of QDs that can interact with the nanostructure. The field intensity influences the absorption of the excitation laser, and consequently the emission of the QW and QDs, as well as the energy transfer between them. The field intensity maps at the acceptor emission wavelength show similar trends, as seen in Figure 2 (g,h,i) and Figure S4.

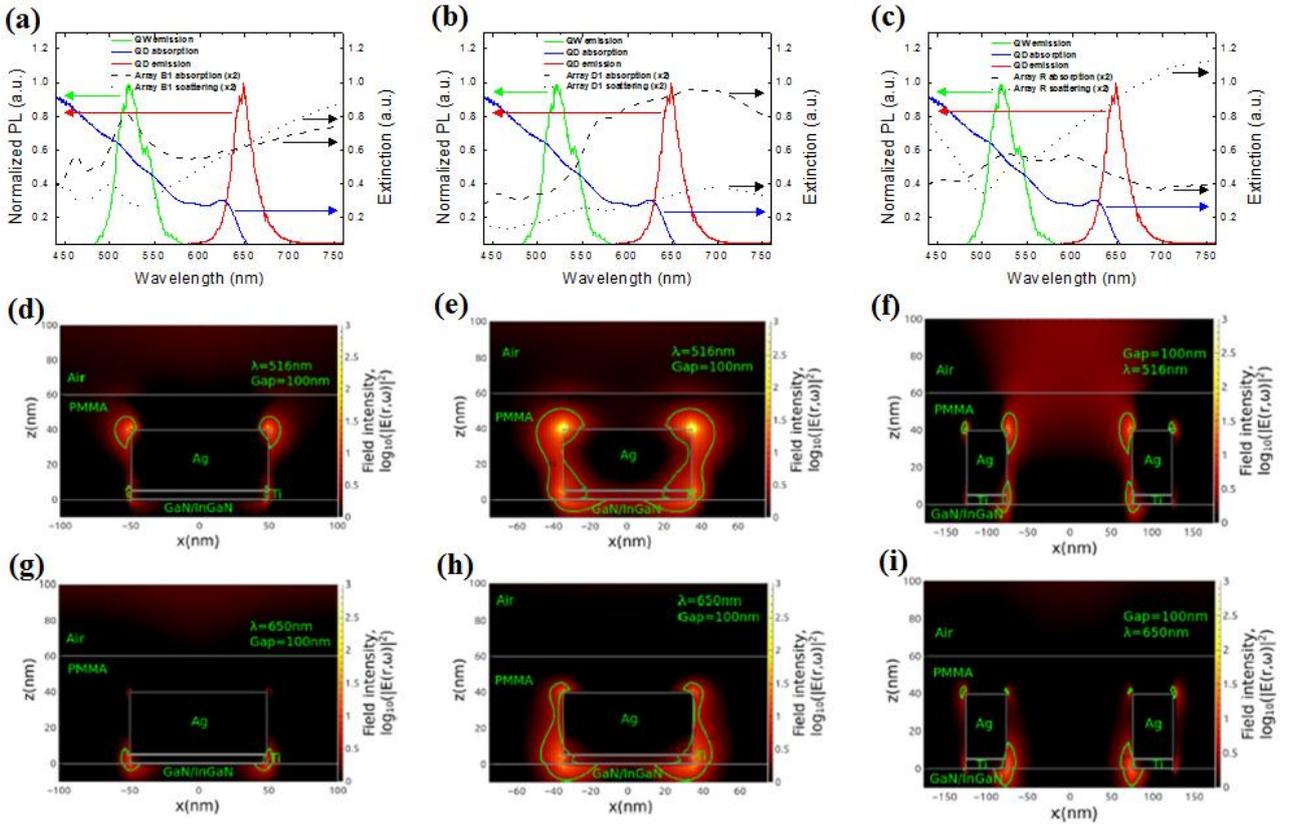

Figure 2 QW PL (green line), QD absorption (blue line) and PL (red line) spectra, along with absorption (dashed black line) and scattering (dotted black line) spectra for arrays B1, D1, R (a,b,c), respectively. Maps of the field intensity in the x-z plane at the donor emission wavelength (d,e,f) and acceptor emission wavelength (g,h,i) for arrays B1, D1 and R, respectively under plane wave excitation. The green line gives $\log_{10} |E|^2 = 0.5$ for direct comparison.

## RESULTS AND DISCUSSION

All measurements are taken at a QW carrier density of $5.4 \times 10^{18}$ cm$^{-3}$ corresponding to the radiative recombination regime for the donor QW. The equation used to calculate the carrier density within the donor QW is presented in the Supporting Information, Equation (S2). The QW PL emission and average PL decay lifetime as a function of carrier density are shown in Figure S5 (a). All measurements were taken at room temperature. The spontaneous emission rate of the donor QW increases with the carrier density in the range above $1 \times 10^{18}$ cm$^{-3}$ and this increasing spontaneous emission rate combined with an $n^2$ dependence of the QW emission on the carrier density, indicates that radiative recombination dominates. The carrier density dependence of plasmon-enhanced FRET has been previously reported.[63] After the deposition of the QD layer on the QW, in the absence of the arrays, there is no measurable effect on the QW lifetime, shown in Figure S5 (b). This demonstrates that there is no significant direct FRET from the QW to the QDs away from the arrays. To facilitate direct comparison of measurements on structures with and without the QDs, the reference undecorated QW and NP decorated QW structures are also covered by a PMMA layer.

To fully characterise plasmon-mediated nonradiative energy transfer the influence of the NP array on the emission properties of the QW and QDs must also be investigated. As an example, the data for the nanoring array will be presented, with similar measurements performed across the seven nanostructure arrays. The analysis will subsequently consider all structures.

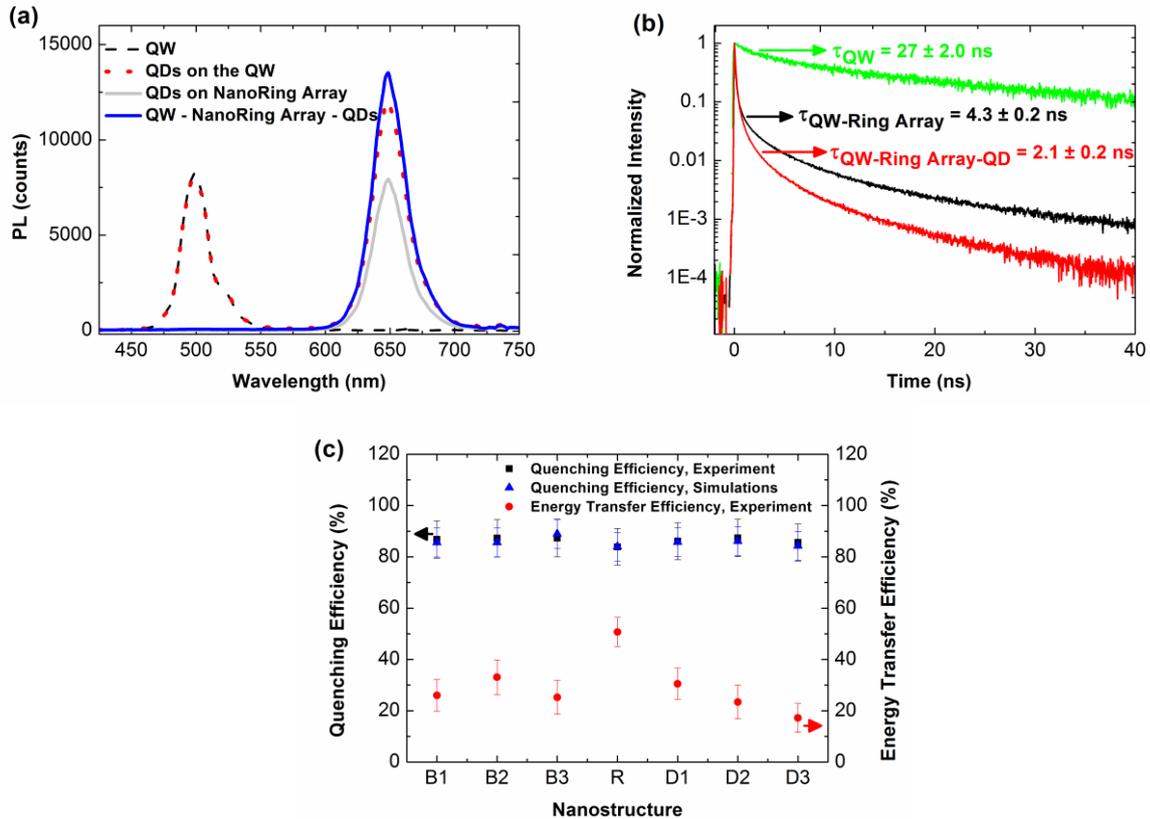

Figure 3: (a) PL spectra of the QW (black dashed line), the QD-QW structure (red dots), the QDs on the nanoring decorated GaN substrate (grey solid line) and the QDs on the nanoring decorated QW substrate (blue solid line). (b) Normalized decay curves of QW only (green line), QW with nanoring array (black line) and QW with nanoring array and QD layer (red line), at a carrier density of $5.4 \times 10^{18}$ cm$^{-3}$. (c) Measured (black squares) and simulated (blue triangles) quenching efficiencies of the donor quantum well and the measured energy transfer efficiency (red circles) as a function of nanostructure.

Figure 3 (a) shows the PL spectra of the QW, the QW with the QDs without the nanoring array, the QDs on nanoring arrays decorated bulk GaN and the array decorated QW with the QDs. From the black and red lines it can be seen that there is no evidence for energy transfer from the QW to the QDs in the absence of the array, as the emission from the QW is the same in both cases. The QW emission is quenched dramatically on the NP array, and the QDs are also quite significantly quenched to approximately (66 ± 5) %. The QW emission quenching is also seen in the QW lifetime data, shown in Figure 3 (b), with a reduction of the QW PL lifetime from (27 ± 2) ns to (4.3 ± 0.2) ns. With the addition of the QD layer there is a further reduction of the QW lifetime to (2.1 ± 0.2) ns, indicating plasmon-enhanced FRET from the QW to the QDs. On the array decorated QW, the QD emission recovers and subsequently, shows an overall QD emission enhancement of approximately (14 ± 1) %.

The quenching efficiency, $E_Q$, is quantified from the PL decay lifetimes of the QW in the absence, $\tau_{QW}$, and presence, $\tau_{QW-NP}$, of the NP array, given for completeness in Equation (1).[10]

$$E_Q = 1 - \frac{\tau_{QW-NP}}{\tau_{QW}} \qquad (1)$$

The quenching efficiency of (84 ± 10) % indicates a very strong interaction between the QW and the nanoring array. A similar quenching efficiency is observed for all seven arrays, with an average of (85 ± 12) %, as can be seen in Figure 3 (c).

To numerically model this quenching process, the theoretical quenching efficiency of the QW is introduced as:

$$E_Q = \frac{k_{QW-NP}^{ET}}{k_{QW-NP}} \qquad (2)$$

where $k_{QW-NP}^{ET}$ represents the energy transfer rate from the donor QW to the acceptor NP array and $k_{QW-NP}$ represents the total decay rate of the Ag array decorated donor QW. Here $k_{QW-NP}^{ET} = k_{QW-NP} - k_{QW}$, where $k_{QW} = \tau_{QW}^{-1}$ is the spontaneous emission rate of the undecorated QW. To simulate the quenching efficiency of the donor QW, a large number (500) of electric point dipoles with random positions and orientations are placed in the InGaN layer representing the donor QW and these dipoles are restricted to the central unit cell of a 5 x 5 array of acceptor NPs. The nonradiative energy transfer and total decay rates of these dipoles are calculated and used to calculate the quenching efficiency from Equation (2). The mean of this distribution is taken as the value of the quenching efficiency, and its standard error generates the error bar. The simulated data for all seven arrays is also shown in Figure 3 (c) and excellent agreement with the experimental data is obtained.

The QW lifetime is further reduced when the QD layer is added, as seen in Figure 3 (b) due to plasmon-enhanced FRET from the QW to the QDs. The efficiency, $E_{FRET}$ which is again experimentally determined from the PL decay lifetimes of the NP array decorated QW in the absence, $\tau_{QW-NP}$, and presence, $\tau_{QW-NP-QD}$, of the QD layer, is given in Equation (3).

$$E_{FRET} = 1 - \frac{\tau_{QW-NP-QD}}{\tau_{QW-NP}} \qquad (3)$$

For the nanoring array the plasmon enhanced energy transfer efficiency is found to be (51 ± 7) %. This represents a significant enhancement of the energy transfer efficiency. As commented earlier, there is no directly measureable FRET from the QW to the QDs when the NP array is not present, seen in Figure S5 (b). The energy transfer is sufficiently enhanced to compete with the very efficient quenching of the QW by the NP array and the intrinsic spontaneous emission rate of the QW.

$E_{FRET}$ for all seven arrays is shown in Figure 3 (c). Whilst $E_Q$ is observed to be similar for all array geometries, the $E_{FRET}$ shows much greater variation. Looking firstly at the three nanobox arrays, the $E_{FRET}$ is approximately 30 % in all cases, and appears to be relatively insensitive to the change in gap from 100 nm (B1) to 160 nm (B3), respectively. Spectral overlap exists between a peak in the absorption spectra of the nanobox arrays and the QW emission spectrum, with the level of absorption decreasing from B1 to B3, as seen in Figure S2 (a,c,e). Despite this clear reduction in the absorption at the donor emission wavelength, $E_{FRET}$ is static amongst the box arrays. On the other hand, the disc arrays show greater sensitivity to the array pitch. $E_{FRET}$ decreases from (30 ± 6) % to (17 ± 6) % as the gap between disc increases from 100 nm (D1) to 160 nm (D3), respectively. This decrease in the LSP-FRET efficiency may be influenced by the red-shift in the absorption spectra with increasing disc spacing from D1 to D3, as shown in Figure S2 (b,d,f), which would reduce the spectral overlap between the donor QW emission spectrum and the disc array absorption spectra going from D1 to D3. However, as commented earlier, $E_{FRET}$ did not simply follow the trend of the absorption spectral changes for the nanobox arrays. The ring array also has a peak in the absorption spectrum overlapping with the QW emission. The peak value is similar to Box 2 as seen from Figures S2 (c) and (g), however a much larger energy transfer efficiency (51 ± 7) % is observed. This could be due to the much greater spatial extent of the electric field intensity observed for the ring array.

The trends seen in the LSP-FRET efficiency with varying plasmonic array cannot be trivially predicted by examining the far-field absorption and scattering spectra simulated using plane wave excitation. These spectra can act as a guide in selecting a plasmonic geometry, however a more complex model is required to explain variations in the plasmon-enhanced FRET across the structures. Such a model must take account of the distribution of donor dipoles at different positions under the arrays within the QW and the field intensity distribution generated by plane wave excitation is different to that generated by dipolar excitation, which is sensitive to the position of the dipole source. Such simulations are computationally expensive requiring averaging over very high numbers of QW donor dipoles at many positions to very high numbers of QD acceptor dipoles at many positions.

The total QD acceptor emission is characterised by integrating the TRPL data. The measurement of the QD emission on and off the array on GaN substrate provides insight into the direct influence of the array on the QD emission. The ratio of the emission on the array to off the array yields the array emission factor shown in Figure 4 (a). It can be seen that the QD emission is much more sensitive to NP array properties than the QW, with the ratio varying from ~ 66 % to ~ 141 %. The sensitivity to the gap between individual nanostructures is highest for boxes. The QD emission is enhanced on structures B1 and B2 by ~ 41 % and ~ 20 %, respectively, but for all other structures the QD emission is quenched. The level of scattering at the acceptor emission wavelength (649 nm) decreases as the gap increases from B1 to B3, as seen in Figure S2 (a,c,e), which is in-line with these experimental observations. While for the disc arrays, the level of absorption decreases from D1 to D3, seen in Figure S2 (b,d,f), which could contribute to the trend of increasing QD emission with increasing gap. The ring array produces the highest direct quenching of the QD emission.

The theoretical array emission factor can be expressed as

$$\frac{I_{em}^{QD-NP}}{I_{em}^{QD}} = \frac{I_{abs}^{QD-NP}}{I_{abs}^{QD}} \frac{k_{rad}^{QD-NP}}{k_{rad}^{QD}} \frac{\tau_{QD-NP}}{\tau_{QD}} \qquad (4)$$

where $\frac{I_{em}^{QD-NP}}{I_{em}^{QD}}$ is the QD emission factor on the GaN substrate;

- $\frac{I_{abs}^{QD-NP}}{I_{abs}^{QD}}$ is the absorbed intensity ratio, given by the field intensity factor at the position of the QD acceptor with and without the array at the excitation frequency, $\omega_{exc}$;

- $\frac{k_{rad}^{QD-NP}}{k_{rad}^{QD}}$ is the ratio the QD radiative decay rates;

- $\frac{\tau_{QD-NP}}{\tau_{QD}}$ is the lifetime ratio of the acceptor QDs, which is measured from the experimental TRPL data.

The superscripts QD-NP and QD denote on and off the array, respectively, on the GaN substrate.

To simulate the array emission factor, a large number (500) of random acceptor QD dipoles are dispersed within the PMMA layer and the volume average of the field intensity within a unit cell of height 80 nm is calculated. As can be seen from Figure 4 (a), good agreement with the experimentally measured QD emission ratio is obtained for all arrays, confirming that the intrinsic QD-array interaction can be explained through both enhanced absorption and emission processes.

It is interesting to examine components of Equation 4 to shed light on how each of the parameters is modified as the plasmonic array geometric properties change. $\frac{\tau_{QD-NP}}{\tau_{QD}}$ shows little variation from one array to another, with values from 0.7 – 0.8 for all arrays, as seen in Figure S6 in the Supporting Information. The other parameters are extracted from the model. Figure 4 (b) shows $\frac{I_{abs}^{QD-NP}}{I_{abs}^{QD}}$ has the largest value for array B1 and decreases with increasing spacing for the three nanobox arrays. There is little variation across the three nanodics arrays and almost no enhancement for the nanoring. The radiative decay ratio for the acceptor QDs, $\frac{k_{rad}^{QD-NP}}{k_{rad}^{QD}}$, shows enhancement of the radiative rate for the nanobox arrays and is again greatest for B1, seen Figure 4 (c). The nanoring and nanodisc arrays all show suppression of the QD radiative rate. The trends of $\frac{I_{abs}^{QD-NP}}{I_{abs}^{QD}}$ and $\frac{k_{rad}^{QD-NP}}{k_{rad}^{QD}}$ are similar to the general trend of the QD emission ratios on GaN.

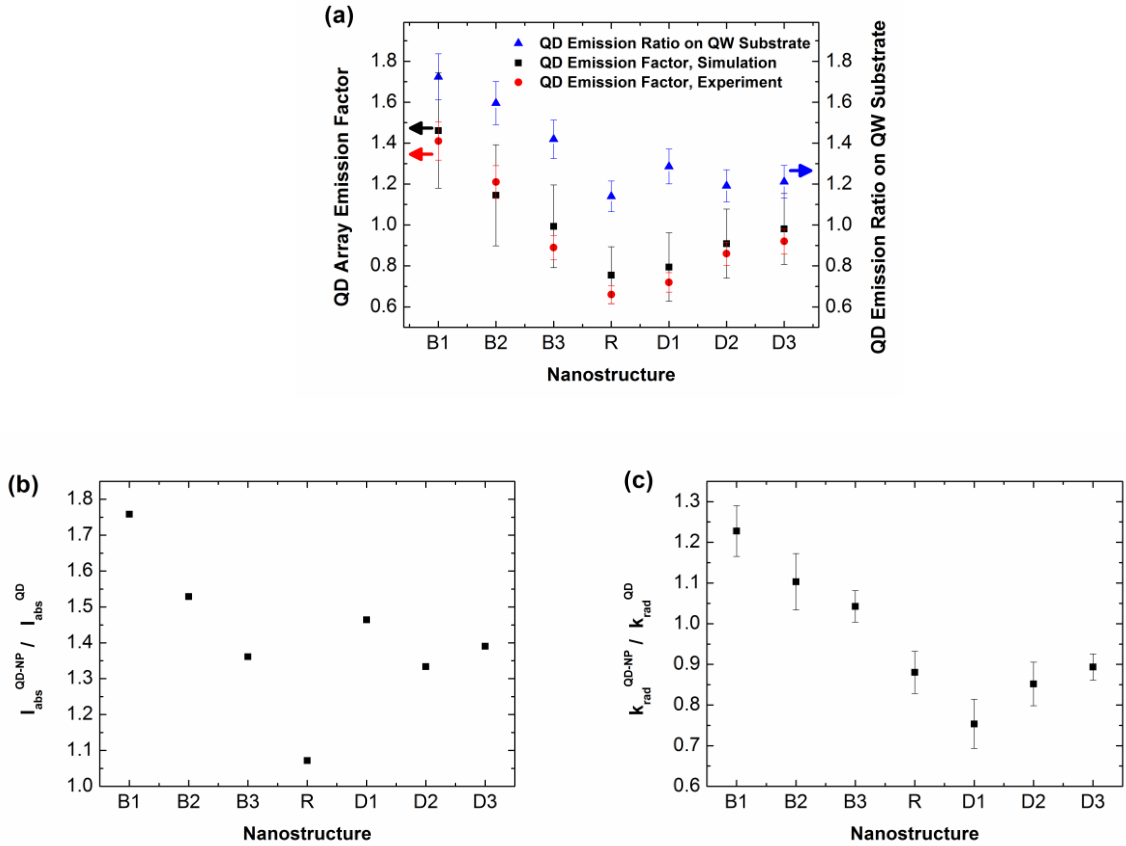

Figure 4: (a) Experimental QD array emission factors (black squares) and theoretical QD array emission factors (red discs) on bulk GaN along with QD emission ratios on QW substrate (blue triangles) as a function of nanostructure. (b) Simulated QD absorbed intensity ratio and (c) simulated QD radiative decay ratio as a function of nanostructure.

The QD emission ratio on the array decorated QW samples is also plotted in Figure 4 (a), allowing direct comparison with the array emission factor on GaN, and is seen to be higher for all arrays. The smallest overall enhancement of QD emission on the QW is observed for the nanoring geometry (~14 %), even though this geometry shows the highest energy transfer efficiency. Box1 shows a further increase relative to the array emission factor and yields the highest overall increase of the QD emission. The QD emission ratio on the QW remains sensitive to the variation in gap dimension across the box arrays but is seen to be static at ~ 1.2 for the three disc arrays. The increased emission on the QW is attributed to plasmon-enhanced FRET. The QD PL decays also provided evidence of pumping of the QDs by the plasmon-enhanced FRET, presented in Figure S7 in the Supporting Information. The average QD lifetime is seen to increase on the array decorated QW, whereas it is reduced on the array decorated GaN substrate. The trend of the change in the QD lifetime across the arrays is the similar to the trend of the plasmon-enhanced FRET efficiency, as seen in Figure S7 (c).

To confirm that the increased QD emission is indeed solely due to plasmon-enhanced FRET from the QW to the QD layer, we model the change in QD emission using the ET efficiency determined experimentally from the QW lifetime data. The theoretical QD emission ratio on the QW can be expressed as

$$\frac{I_{em}^{QD-NP-QW}}{I_{em}^{QD}} = \frac{I_{em}^{QD-NP}}{I_{em}^{QD}} \left[ 1 + \frac{A_{QW}(\omega_{exc})}{A_{QD}(\omega_{exc})} \frac{I_{abs}^{QW}}{I_{abs}^{QD}} E_{FRET} \right] \quad (5)$$

where the first term, $\frac{I_{em}^{QD-NP}}{I_{em}^{QD}}$, corresponds to the array emission factor, given by Equation (4), and the second term, $1 + \frac{A_{QW}(\omega_{exc})}{A_{QD}(\omega_{exc})} \frac{I_{abs}^{QW}}{I_{abs}^{QD}} E_{FRET}$, can be considered as the energy transfer factor in which

- $\frac{A_{QW}(\omega_{exc})}{A_{QD}(\omega_{exc})}$ is the ratio of the QW and QD field intensity factors;

- $\frac{I_{abs}^{QW}}{I_{abs}^{QD}}$ is the ratio of the intrinsic absorption in the donor QW to the intrinsic absorption in the acceptor QDs, in the absence of the array, taken from experiment;

- $E_{FRET}$ is the plasmon-mediated energy transfer efficiency determined experimentally from the QW TRPL data using Equation (3) and plotted in Figure 3(c).

The energy transfer factor can be experimentally isolated by plotting the ratio of the QD emission ratio on the QW and on the GaN substrate, corresponding to $\frac{I_{em}^{QD-NP-QW}}{I_{em}^{QD-NP}}$, presented in Figure 5 (a). In the absence of any energy transfer the energy transfer factor is equal to 1.

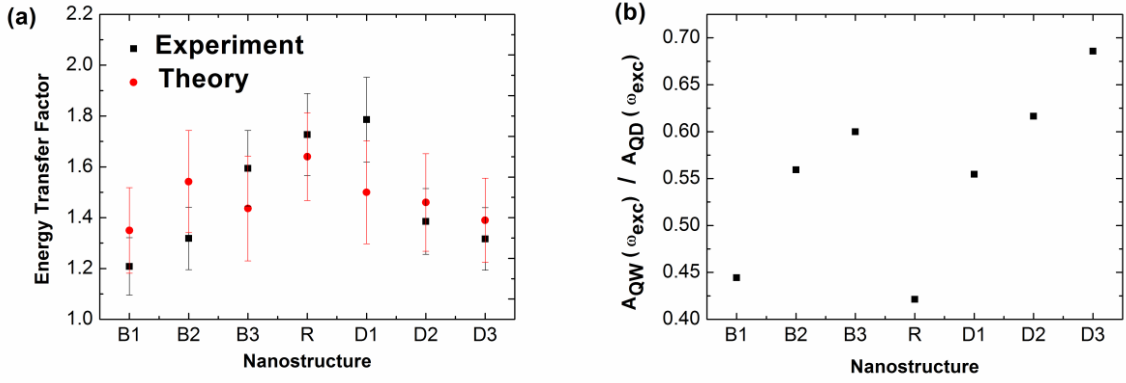

Figure 5: (a) Experimental (black squares) and theoretical (red circles) energy transfer factors as a function of nanostructure. (b) Ratio of the QW and QD field intensity factors at the excitation laser frequency, $\omega_{exc}$.

Good agreement between the experimental and theoretical results is obtained for all arrays, confirming that the total change in the QD emission from the GaN substrate to the QW substrate is exclusively due to the plasmon-enhanced FRET from the QW to the QDs. This study shows the strong sensitivity of both the array emission factor and the plasmon-enhanced energy transfer factor to the properties of the plasmonic arrays. Comparison of Figures 4 (a) and 5 (a) emphasises the consequence of the trade-off between the direct influence of the array on the QD emission and plasmon-enhanced FRET. The nanoring structure is the best for energy transfer and facilitates the largest increase in the QD emission due the plasmon-enhanced FRET but this is strongly offset by the large direct quenching of the QD emission by the nanoring array. The nanobox array (B1) provides the largest total emission of the QDs for any metal structures studied, by combining smaller gains due to energy transfer with a larger emission enhancement factor.

Again it is interesting to see how the parameters contribute to the change in the QD intensity. $\frac{I_{abs}^{QW}}{I_{abs}^{QD}}$ is constant as it is independent of the NP array. $\frac{A_{QW}(\omega_{exc})}{A_{QD}(\omega_{exc})}$, is the ratio of the field intensity factors for the QW and the QD layer, shown in Figure 5 (b) and is found to have a value less than 1 for all arrays, with the ratio rising slightly with increasing gap between individual components for both the disc and the box arrays. It is lowest for the ring array, reducing the benefit of the large energy transfer efficiency.

## CONCLUSION

In conclusion, plasmon-enhanced FRET from an InGaN/GaN QW to a ~ 80 nm layer of QDs embedded in PMMA has been demonstrated using seven different silver NP arrays fabricated by HIL. Arrays with different nanoparticle shape, size and gap were investigated. The GaN barrier was fixed at 3 nm to enable strong near-field coupling between the donor QW and the plasmonic arrays. The arrays were fabricated directly on QW and GaN substrates. The modification of the QD emission on the array decorated GaN substrates is shown to be due to enhanced absorption and enhanced radiative decay, with the nanobox arrays showing the potential for greater enhancement and tunability. The energy transfer efficiency is shown to be highly sensitive to the array properties, varying from 25 % to 51 %, with the nanoring array yielding the highest observed efficiency. The efficiency is relatively insensitive to the variation in gap size for the nanobox arrays, and shows greater variability for the nanodisc arrays. As there is no measureable direct FRET from the QW to the QDs, the plasmonic arrays are seen to facilitate large increases in the FRET efficiency, despite the strong quenching of the QW emission. The energy transfer process also manifested as longer QD decay lifetimes, with increases of up to 60 %. The QD emission is increased on the array decorated QW substrates compared with the GaN substrate for all arrays, and by comparison with simulation results it is shown that this increase can be attributed exclusively to plasmon-enhanced FRET. The energy transfer contribution to the QD emission is again highly sensitive to the array geometry and the energy transfer factor can be as high as 1.7 for the nanoring array. However, this competes with the strong quenching influence of the array on the QD emission to yield just a ~14 % enhancement of the QD emission on the QW. QD emission enhancements of up to ~ 71 % were observed for a nanobox array which combined more modest gains due to the energy transfer with increased QD excitation and radiative rates. Arising from the trade-offs between the different parameters determining the QD emission, the plasmon enhanced system is relatively insensitive to the gap spacing for the nanodisc arrays and shows greater variability for the nanoboxes.

These results show tunability of the plasmonic arrays for energy transfer or colour tuning applications. Structures can be designed to optimise FRET, or to benefit from combined ET and enhanced emission mechanisms for light emitter applications. The trends seen in the plasmon-enhanced FRET efficiency with varying plasmonic array cannot be simply predicted by examining the simulated far-field absorption and scattering spectra alone. A more complex model is required, in which the distribution of donor and acceptor dipoles is taken into account. It should also be noted that for future device applications the QW can be electrically pumped, not optically pumped as we have demonstrated here. The increase in the energy efficiency could allow for a thicker GaN capping layer, alleviating carrier transport problems associated with thin GaN capping layers.

**ACKNOWLEDGEMETS**

This work was supported by Science Foundation Ireland (SFI) under grant number 10/IN.1/I2975 and the National Access Programme Grant under grant number NAP 338, and enabled using facilities funded by Irish Higher Education Authority Programme for Research in Third Level Institutions Cycles 4 and 5 via the INSPIRE and TYFFANI projects. We also acknowledge the Advanced Microscopy Laboratory (AML) for use of the helium-ion microscope. JJG acknowledges a postgraduate research scholarship from the Irish Research Council (GOIPG/2013/680) and PJP a SFI Engineering Professorship (SFI/07/ EN/E001A).